\documentclass[fleqn,usenatbib]{mnras}

\usepackage{newtxtext,newtxmath}
\usepackage[T1]{fontenc}
\usepackage{ae,aecompl} 

\usepackage{graphicx}	% Including figure files
\usepackage{amsmath}	% Advanced maths commands
\usepackage{amssymb}	% Extra maths symbols
\usepackage{color}	% Extra maths symbols

\usepackage{amsmath}%Loaded by Oscar
\usepackage{breqn}%Loaded by Oscar
\usepackage{layouts}%Loaded by Oscar, to measure textwidth

\title[Turbulence profiles at San Pedro M\'artir]{Simultaneous Generalized and Low-Layer SCIDAR turbulence profiles at San Pedro M\'artir Observatory}

\author[R.~Avila et al.]{
R.~Avila,$^{1}$\thanks{E-mail: remy@fata.unam.mx}
O.~Vald\'es-Hern\'andez,$^{1}$
L.~J.~S\'anchez,$^{2}$
I.~Cruz-Gonz\'alez,$^{2}$
\newauthor
J.~L.~Avil\'es,$^{3}$
J.~J.~Tapia-Rodr\'iguez,$^{4}$
and C.~A.~Z\'u\~niga$^{1}$
\\
$^{1}$Centro de F\'isica Aplicada y Tecnolog\'ia Avanzada, Universidad Nacional Aut\'onoma de M\'exico, A.P. 1-1010, Quer\'etaro,  ~~Qro.  76000, M\'exico\\
$^{2}$Instituto de Astronom\'ia, Universidad Nacional Aut\'onoma de M\'exico, A.P. 70-264, Ciudad Universitaria 04510, Ciudad de ~~M\'exico,  M\'exico\\
$^{3}$Tecnol\'ogico Nacional de M\'exico, Dpto. de Ciencias B\'asicas, Av. Tecnol\'ogico s/n Esq. M. Escobedo, Centro, Quer\'etaro, Qro. 76000, M\'exico\\ 
$^{4}$Instituto Tecnol\'ogico de Morelia, Avenida Tecnol\'ogico 1500, Lomas de Santiaguito, Morelia, Michoac\'an Mich., 58120, M\'exico
}

\date{Accepted XXX. Received YYY; in original form ZZZ}

\pubyear{2019}

\begin{document}
\label{firstpage}
\pagerange{\pageref{firstpage}--\pageref{lastpage}}
\maketitle

\begin{abstract}

We present optical turbulence profiles obtained with a Generalized SCIDAR (G-SCIDAR) and a Low Layer SCIDAR (LOLAS) at the Observatorio Astron\'omico Nacional in San Pedro M\'artir (OAN-SPM), Baja California, Mexico, during three observing campaigns in 2013, 2014 and 2015. The G-SCIDAR delivers profiles with moderate altitude-resolution (a few hundred meters) along the entire turbulent section of the atmosphere, while the LOLAS gives high altitude-resolution (on the order of tens of meters) but only within the first few hundred meters. Simultaneous measurements were obtained on 2014 and allowed us to characterize in detail the combined effect of the local orography and wind direction on the turbulence distribution close to the ground. At the beginning of several nights, the LOLAS profiles show that turbulence peaks between 25 and 50 m above the ground, not at ground level as was expected.
The G-SCIDAR profiles exhibit a peak within the first kilometer. In 55\% and 36\% of the nights  stable layers are detected between 10 and 15 km and at 3 km, respectively.
This distribution is consistent with the results obtained with a G-SCIDAR in 1997 and 2000 observing campaigns. Statistics computed with the 7891 profiles that have been measured at the OAN-SPM with a G-SCIDAR in 1997, 2000, 2014 and 2015 campaigns are presented. The seeing values calculated with each of those profiles have a median of 0.79, first and third quartiles of 0.51 and 1.08 arcsec, which are in close agreement with other long term seeing monitoring performed at the OAN-SPM.

\end{abstract}


%%%%%%%%%%%%%%%%% BODY OF PAPER %%%%%%%%%%%%%%%%%%

\section{Introduction}
\label{sec:intro}

In optical astronomy, the Earth's atmosphere is frequently the main cause of angular resolution degradation when using large telescopes. It is a direct effect of turbulent fluctuations of the refractive index along the light path. Those fluctuations are commonly referred to as optical turbulence. Adaptive optical (AO) systems are intended to correct for the phase deformations introduced by optical turbulence. Huge efforts are dedicated to the development of modern AO systems in new generation telescopes. As telescopes become larger, the performance of AO systems becomes more sensitive to variations of the vertical distribution of the optical turbulence strength $C_N^2(h)$ \citep{doi:10.1093/mnras/stw2685,Neichel:09,Basden:10,Vidal:10,MRF+13}. 

The characterization of the turbulence profiles $C_N^2(h)$ at astronomical sites where new telescopes are to be installed is of crucial importance for the optimization of the instrumental performance and the quality of the scientific return.

A number of optical turbulence profiling techniques have been developed. Among those, the most prevalent that are in operation today are SCIDAR (Scintillation Detection And Ranging, \citet{1973JOSA...63..270V}) SLODAR (Slope Detection and Ranging, \citet{2002MNRAS.337..103W}) and Multi Aperture Scintillation {\color{black}Sensor} (MASS, \citet{TK07}). SCIDAR and SLODAR are based on a triangulation method, both use a double star of known separation as light source, while MASS uses a single star. MASS has a fixed and coarse altitude resolution of the turbulence profile, while with SCIDAR and SLODAR, the altitude resolution depends on the double star angular separation. The original SCIDAR technique was insensitive to turbulence at altitudes lower than 1~km. It then evolved to the Generalized SCIDAR (G-SCIDAR, \citet{FTV98,AVM97}) that overcame that limitation. The G-SCIDAR further branched into the Low Layer SCIDAR (LOLAS, \citet{2008MNRAS.387.1511A}), for high resolution profiles close to the ground, and to the Stereo-SCIDAR \citep{2014MNRAS.437.3568S}, for ultimate sensitivity. 

Instruments of G-SCIDAR and LOLAS methods have been developed at the Universidad Nacional Aut\'onoma de M\'exico (UNAM) and reported in \citet{2003RMxAC..19...44C} and \citet{2012MNRAS.423..900A, 2016PASP..128j5001A}, respectively. Here we report results obtained with these two instruments at the Observatorio Astron\'omico Nacional in San Pedro M\'artir (OAN-SPM), Baja California, Mexico during three campaigns in 2013, 2014 and 2015. Turbulence and wind profiles measured at the same site with the G-SCIDAR developed at the University of Nice have been reported by \citet{1998PASP..110.1106A,2004PASP..116..682A,2006PASP..118..503A,2007RMxAC..31...71A}.

%%%%%%%%%%%%%%%%%%%%%%%%%%%%%%%%%%%%%%%%%%%%%%%%%%

\section{Methods}
\label{sec:method}

The SCIDAR method and its derivatives have been extensively explained in the literature. In this section we give an overview and emphasize aspects particularly relevant to this paper. 

%%%%%%%%%%%%%%%%%%%%%%%%%%%%%%%%%%%%%%%%%%%%%%%%%%

\subsection{SCIDAR}

Refractive-index variations in a turbulent layer produce phase fluctuations on light waves coming from a star, which upon propagation down to the ground develop scintillation patterns that can be measured on the pupil plane of a telescope. When observing a double star of known angular separation $\vec{\theta}$,
the speckle patterns produced by each star are identical but shifted a distance $\vec{\theta}z$ from each other, $z$ being the distance from the pupil plane to the turbulent layer along the optical axis. By measuring this distance, $z$ can be retrieved. This {\color{black}is not} done on a single scintillation image because it is constituted by a multitude of superimposed speckles that fill the pupil. Instead, the spatial autocovariance of the composite scintillation is computed from thousands of scintillation images whose exposure time is short enough to freeze the speckles. The autocovariance exhibits two peaks at positions $\vec{r} = \pm \vec{\theta} z$ with an amplitude proportional to the refractive-index structure constant $C_N^2$ of that layer \citep{1974JOSA...64.1000R}. The determination of the position and amplitude of those peaks leads to $C_N^2(z)$. This is the principle of the so-called classical SCIDAR, in which the scintillation is recorded at ground level by taking images of the telescope pupil. 
Since the turbulence of different layers is statistically independent from each other, the overall scintillation autocovariance is the sum of the contribution of each layer and can be written as \citep{AVS01}:   
\begin{equation}
\Lambda(\vec{r})=\int_0^\infty \mathrm{d}z\; C_N^2(z) \left\{aB(\vec{r},z) + b\left[B(\vec{r}+\vec{\theta} z,z)+B(\vec{r}-\vec{\theta} z,z)\right]\right\},
\label{Autocov}
\end{equation}
where $B(\vec{r},z)$ represents the theoretical single-star scintillation autocovariance produced by a layer at distance $z$; while  $a$ and $b$ are constants that depend on the magnitude difference of the double star components. Equation~\ref{Autocov} can be decomposed as the sum of one central peak $\Lambda_c(\vec{r})$ where the contribution of all layers accumulates, and two symmetrical collections of lateral peaks $\Lambda_l(\vec{l})$ and $\Lambda_r(\vec{r})$:

%%%%%%%%%%%%%%%%%%%%%%%%%%%%%%%%%%%%%%%%%%%%%%%%%%

\begin{eqnarray}
\label{Autocov-dec-1}
\Lambda(\vec{r})=\Lambda_c(\vec{r})+\Lambda_l(\vec{r})+\Lambda_r(\vec{r}),\\
\label{Autocov-dec-2}
\Lambda_c(\vec{r})=\int_0^\infty \mathrm{d}z\; a\,C_N^2(z) B(\vec{r},z),\\
\label{Autocov-dec-3}
\Lambda_l(\vec{r})=\int_0^\infty \mathrm{d}z\; b\,C_N^2(z)B(\vec{r}+\vec{\theta} z,z),\\
\label{Autocov-dec-4}
\Lambda_r(\vec{r})=\int_0^\infty \mathrm{d}z\; b\,C_N^2(z)B(\vec{r}-\vec{\theta} z,z).
\end{eqnarray}

All the information needed to retrieve $C_N^2(z)$ is contained in either of the lateral terms, for example Eq.~\ref{Autocov-dec-4}, where
\begin{equation}
    b\,=\,\frac{\gamma}{(1+\gamma)^2},
    \label{eq:b}
\end{equation}
with $\gamma=10^{-0.4\Delta m}$, and $\Delta m$ being the stellar magnitude difference. 

%%%%%%%%%%%%%%%%%%%%%%%%%%%%%%%%%%%%%%%%%%%%%%%%%%

\begin{figure}
\centering
\includegraphics[width=\columnwidth]{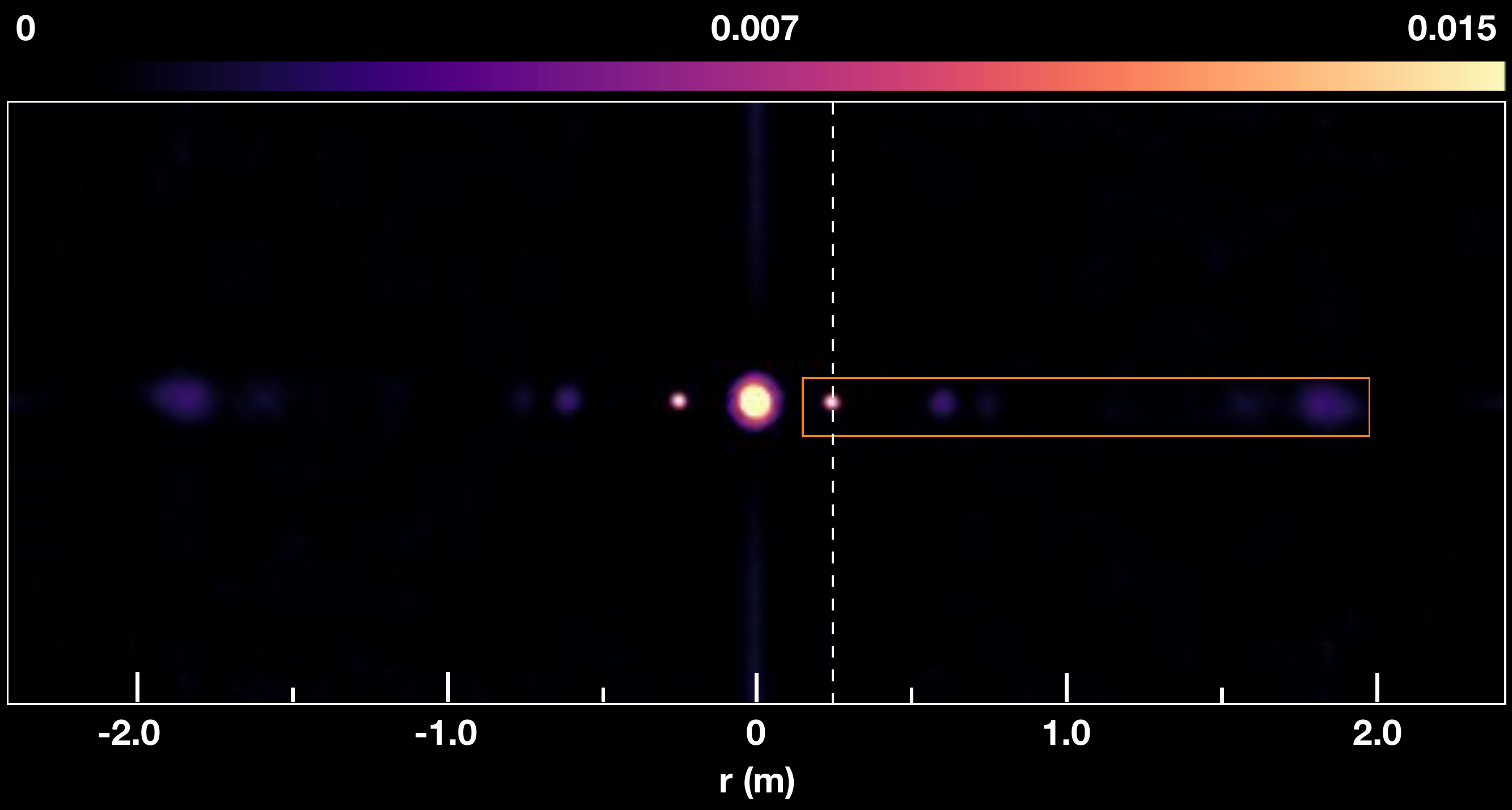}
\includegraphics[width=\columnwidth]{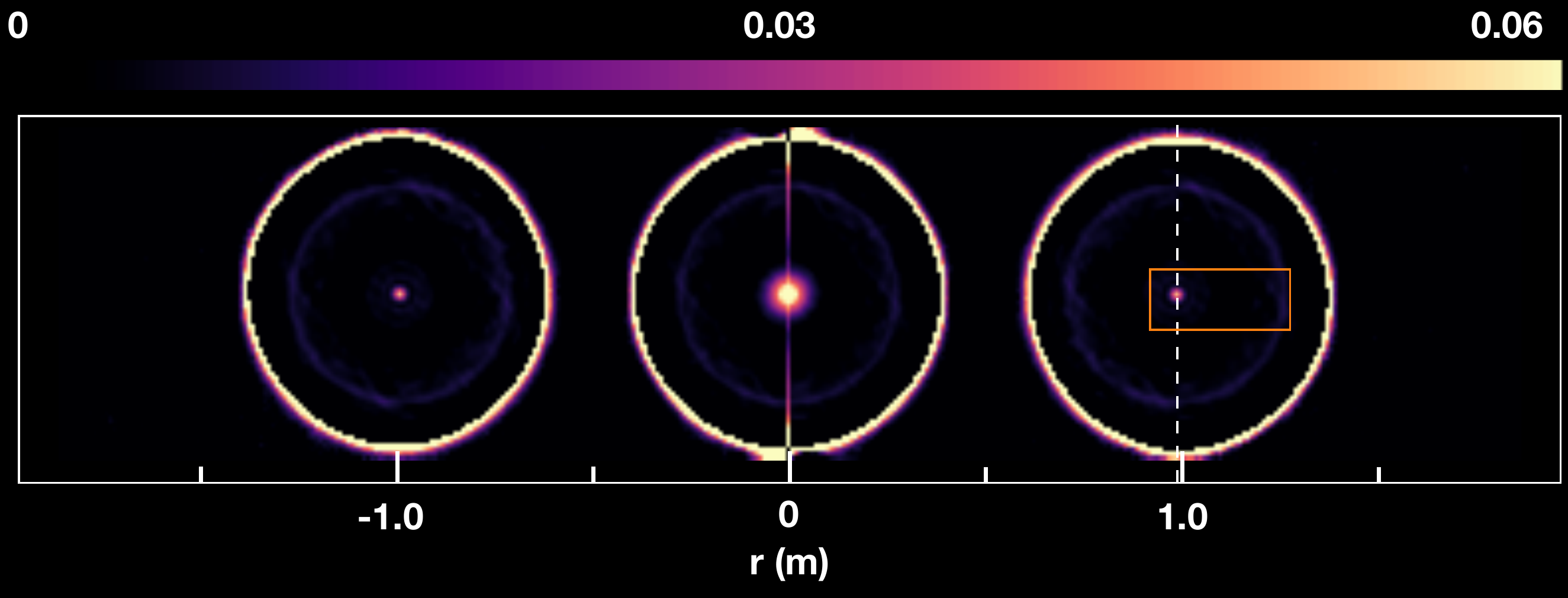}
\caption{Typical scintillation autocovariance maps obtained with the G-SCIDAR (top) and the LOLAS (bottom). Orange rectangles represent the useful zones from where a horizontal cut is taken as the input data for the inversion procedures. The vertical dashed lines indicate the distance $r$ corresponding to ground level, i.e. $r\,=\,{\vec{\theta}} d$. The G-SCIDAR and LOLAS autocovariances shown were obtained at 8:34 on May 07 2015 UT and at 10:41 on November 18 2013, and using stars STF1744AB and 11,\,12 Cam (see Table~\ref{targetsT}), respectively. }
\label{figAcov}
\end{figure}

%%%%%%%%%%%%%%%%%%%%%%%%%%%%%%%%%%%%%%%%%%%%%%%%%%

\subsection{Generalized SCIDAR}

The scintillation variance produced by a turbulent layer at a distance $z$ from the detecting plane is proportional to $z^{5/6}$. Thus, the SCIDAR is blind to turbulence close to the ground, which constitutes a major disadvantage because
the most intense turbulence is often located at ground level, as shown for example by \citet{2004PASP..116..682A}, and \citet{CWA+09}. 
To circumvent this limitation, \citet{FTV94}
proposed to optically shift the measurement plane 
a distance $d$ below the pupil. This is the
principle of the G-SCIDAR which was first implemented by
\citet{AVM97}. For the scintillation variance to be
significant, $d$ must be of the order of 1~km or larger. Another important consequence of adding $d$ to the propagation distance is that the autocovariance lateral peak of the turbulence at ground level gets shifted away from the central peak by a distance $\vec{\theta} d$, making them easily distinguishable from the central peak. An example of scintillation autocovariance obtained with the G-SCIDAR is shown in the top panel of Fig.~\ref{figAcov}. The effective propagation distance in G-SCIDAR becomes the distance from a turbulent layer to the pupil plane $z$ plus the virtual propagation distance $d$. Expressing $z$ in terms of the vertical altitude above the ground $h$ and the angle between the zenith and the direction of the double star $\eta$, the equivalent of Eq.~\ref{Autocov-dec-4} for the G-SCIDAR then reads:

%%%%%%%%%%%%%%%%%%%%%%%%%%%%%%%%%%%%%%%%%%%%%%%%%%

\begin{equation}
    \label{Autocov-GS}
   \Lambda_{r,\mathrm{GS}}(\vec{r})=\int\limits_{\frac{-d}{\sec(\eta)}}^\infty \mathrm{d}H\; b\,C_N^2(h)B\left(\vec{r}-\vec{\theta}H,H\right),\\ 
 \end{equation}
where 
\begin{equation}
    \label{H}
   H=h \sec(\eta)+d.
\end{equation}

%%%%%%%%%%%%%%%%%%%%%%%%%%%%%%%%%%%%%%%%%%%%%%%%%%

\subsection{Low Layer SCIDAR (LOLAS)}
 
\citet{2008MNRAS.387.1511A} introduced the LOLAS as a technique capable of measuring turbulence near ground with high altitude-resolution. It is based on the same concept as the G-SCIDAR method but it uses a much widely separated double star. The typical stellar separation $\vec{\theta}$ for the G-SCIDAR is around $10\arcsec$ while for LOLAS it is larger than $100\arcsec$. Since the maximum attainable altitude $h_{max}$ and the altitude resolution $\delta h$ with SCIDAR techniques are both inversely proportional to $\theta$ \citep{2008MNRAS.387.1511A, 2001A&A...371..366P}: 

%%%%%%%%%%%%%%%%%%%%%%%%%%%%%%%%%%%%%%%%%%%%%%%%%%

\begin{equation}
h_{max}=\frac{D}{{\vec{\theta}}\sec(\eta)},
\label{hmax}
\end{equation}
and
\begin{equation}
\delta h=0.78\frac{\sqrt{\lambda\left(h\sec(\eta)+d\right)}}{\vec{\theta}},
\label{hress}
\end{equation}

%%%%%%%%%%%%%%%%%%%%%%%%%%%%%%%%%%%%%%%%%%%%%%%%%%

 The maximum attainable altitude is drastically reduced in LOLAS, at the gain of a much better altitude resolution. $\lambda$ represents the wavelength of the stellar radiation. $\delta h$ can reach values as low as 10 metres in LOLAS, while the typical SCIDAR resolution is several hundreds of metres. The value of $h_{max}$ gets further reduced because the pupil diameter $D$ in LOLAS is 40~cm, for the instrument to be portable. \citet{Avila:09} have shown that due to the fact that the two defocused pupil images are not superimposed on the detector in LOLAS, the lateral peaks of the scintillation autocovariance are not affected by factor $b$, like in Eqs.~\ref{Autocov-dec-4} and \ref{Autocov-GS}. This is also the case in the Stereo-SCIDAR \citep{2014MNRAS.437.3568S}. The equivalent of Eq.~\ref{Autocov-dec-4} for LOLAS is then simply  
 
%%%%%%%%%%%%%%%%%%%%%%%%%%%%%%%%%%%%%%%%%%%%%%%%%%

\begin{equation}
   \Lambda_{r,\mathrm{LOLAS}}(\vec{r})=\int\limits_{\frac{-d}{\sec(\eta)}}^\infty \mathrm{d}H\; C_N^2(h)B\left(\vec{r}-\vec{\theta}H,H\right).
    \label{Autocov-LOLAS}
\end{equation} 
{\color{black}A} typical example of an autocovariance obtained with the LOLAS is shown in the bottom panel of Fig.~\ref{figAcov}.  

%%%%%%%%%%%%%%%%%%%%%%%%%%%%%%%%%%%%%%%%%%%%%%%%%%
%%%%%%%%%%%%%%%%%%%%%%%%%%%%%%%%%%%%%%%%%%%%%%%%%%

\begin{figure}
\centering
\includegraphics[width=\columnwidth]{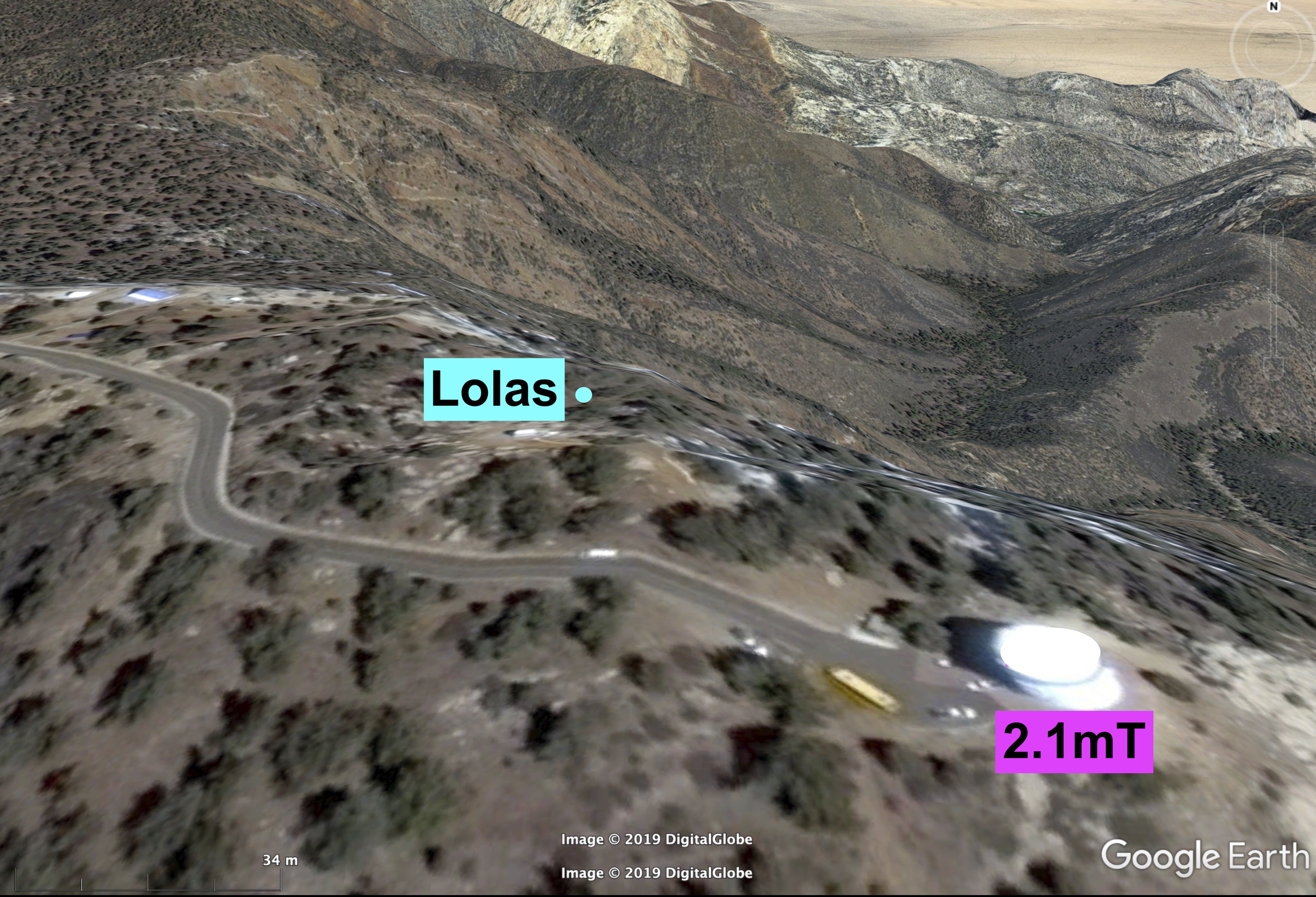}
\includegraphics[width=\columnwidth]{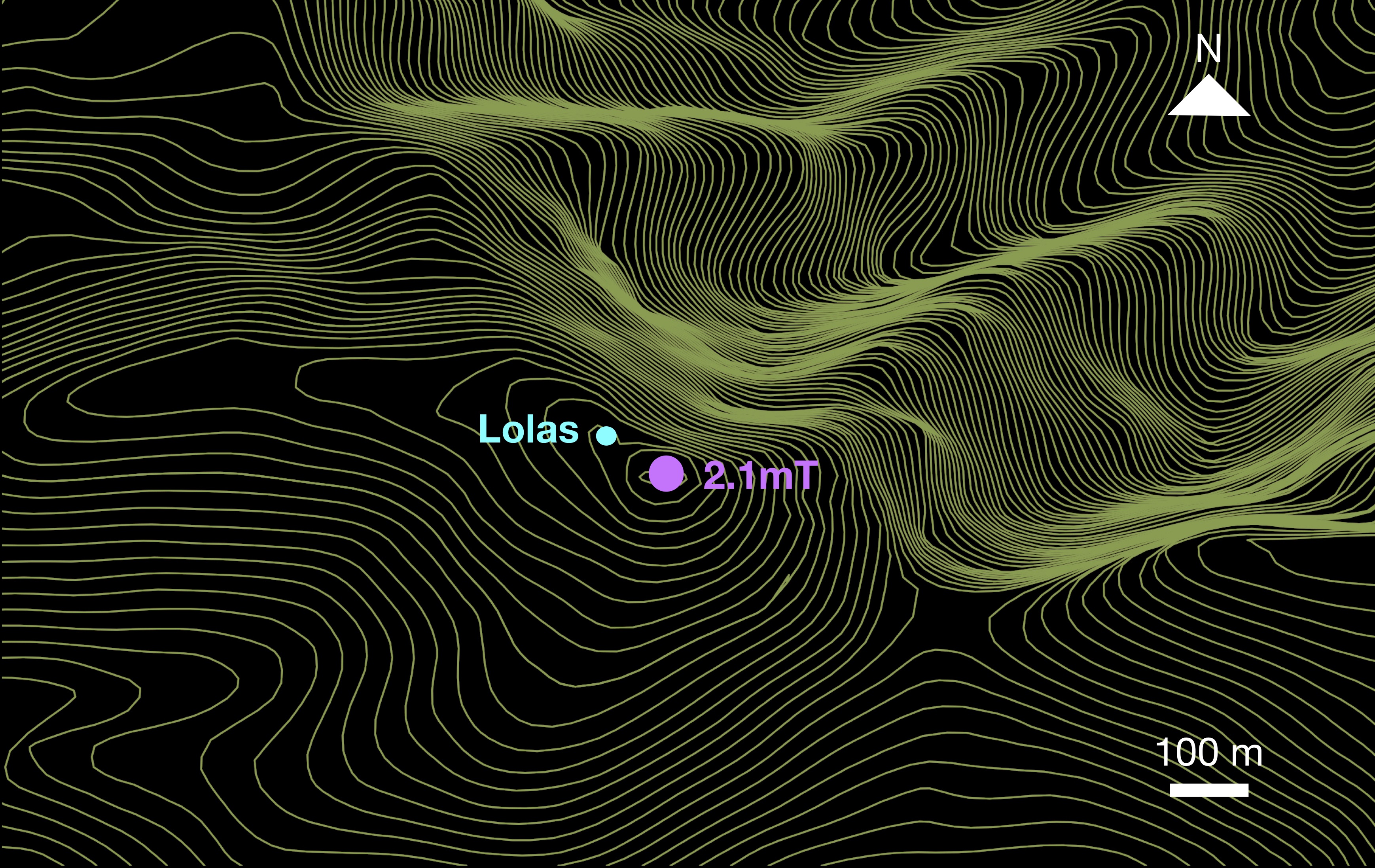}
\caption{
Aerial view (top) of the OAN-SPM summit at San Pedro M\'artir. LOLAS equatorial mount  was fixed to a 1-m tall concrete column on a hill that faced  a  cliff  towards  the  North-East.  The G-SCIDAR was installed at the 2.1-m telescope which has a height of 20 m at the telescope floor. A contour map level curves (bottom) of the site shows the location and separation of the two instruments. Contour curves are separated by 6 m.}
\label{maps}
\end{figure}

%%%%%%%%%%%%%%%%%%%%%%%%%%%%%%%%%%%%%%%%%%%%%%%%%%

\subsection{Obtaining turbulence profiles}

From the Wiener-Khinchin theorem, the autocovariance function can be calculated by the Fourier transform of the power spectrum of the irradiance fluctuations. Assuming Kolmogorov turbulence, \citet{1981PrOpt..19..281R} derived an expression of the scintillation power spectrum of the irradiance fluctuations, which has circular symmetry. Its Fourier transform can then be written as a Hankel transform. For a single star and a unique turbulent layer at a distance $z$ from the analysis plane, the theoretical scintillation autocovariance can be expressed as:

%%%%%%%%%%%%%%%%%%%%%%%%%%%%%%%%%%%%%%%%%%%%%%%%%%

\begin{equation}
B(\vec{r},z)= 0.243 k^2\int_0^\infty \mathrm{d}f f^{-8/3}\sin^2{\left( \pi \lambda z f^2\right)} J_0\left( 2\pi f r\right),
\label{acovsingle}
\end{equation}
%%%%%%%%%%%%%%%%%%%%%%%%%%%%%%%%%%%%%%%%%%%%%%%%%%
where $k=\frac{2\pi}{\lambda}$ and $J_0$ is the first order Bessel function. 
\bigskip
A slice of pixels $\Lambda_r^-$ taken from the center of $\Lambda_{r,\mathrm{GS}}$ or $\Lambda_{r,\mathrm{LOLAS}}$ and along the separation of the stars can be used as a measured vector to solve Eqs.~\ref{Autocov-GS} or \ref{Autocov-LOLAS} by non-negative least squares as suggested by \citet{2014MNRAS.437.3568S}. The linear equation has the form

%%%%%%%%%%%%%%%%%%%%%%%%%%%%%%%%%%%%%%%%%%%%%%%%%%
\begin{equation}
AC_N^2(z)=\Lambda_d^-\;,
\label{linearProb}
\end{equation} 

where $A$ is a matrix whose columns $j$ are the response functions $b\,B(\vec{r}-\vec{\theta}H_j,H_j)$ for G-SCIDAR or $B(\vec{r}-\vec{\theta}H_j,H_j)$ for LOLAS, for each value of $H$.
To avoid over-sampling we solved the least squares problem by using a matrix $A$ which is zero everywhere except in the columns that correspond to multiples of the natural resolution of the instrument given by Eq.~\ref{hress}.

\citet{Avila:09} found that the $C_N^2$ values obtained with the G-SCIDAR method explained so far must be multiplied by a correction factor to be correctly calibrated. This factor depends on ${\vec{\theta}}$, $d$, $D$ and $b$. The G-SCIDAR $C_N^2$ values reported here have been corrected following \citet{Avila:09} and \citet{ASC+11}. In the case of LOLAS, \citet{Avila:09} showed that there is no correction needed. 

%%%%%%%%%%%%%%%%%%%%%%%%%%%%%%%%%%%%%%%%%%%%%%%%%%

\begin{figure}
\centering
\includegraphics[width=\columnwidth]{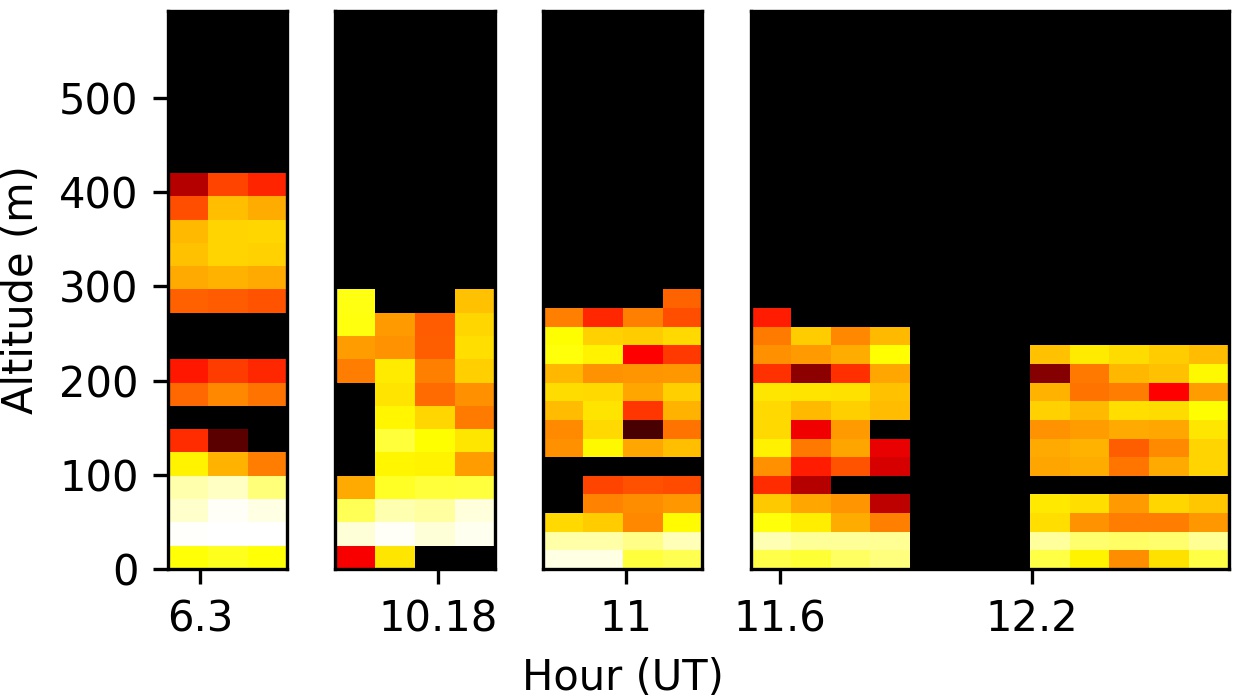} \\
\includegraphics[width=\columnwidth]{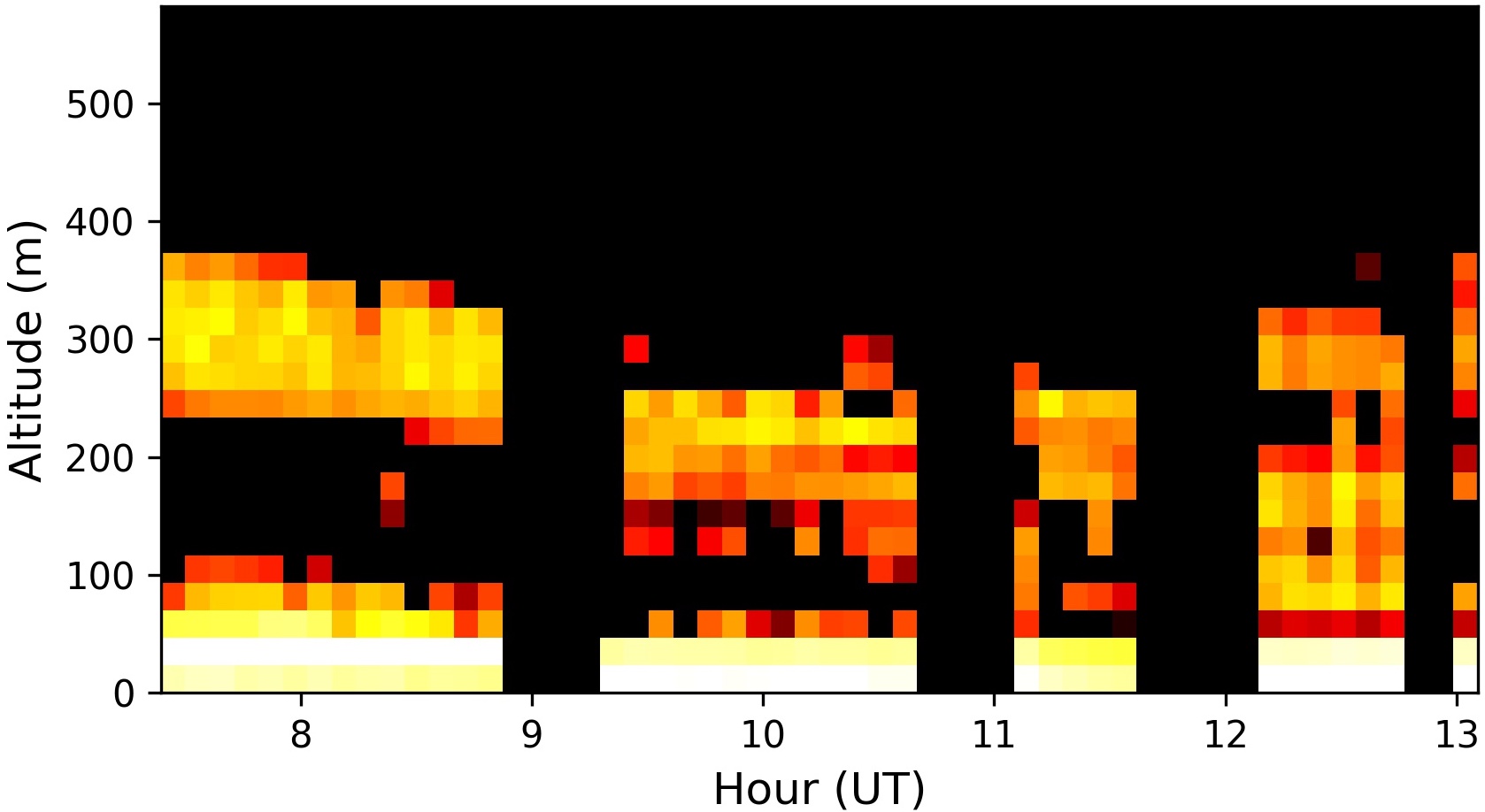} \\\includegraphics[width=\columnwidth]{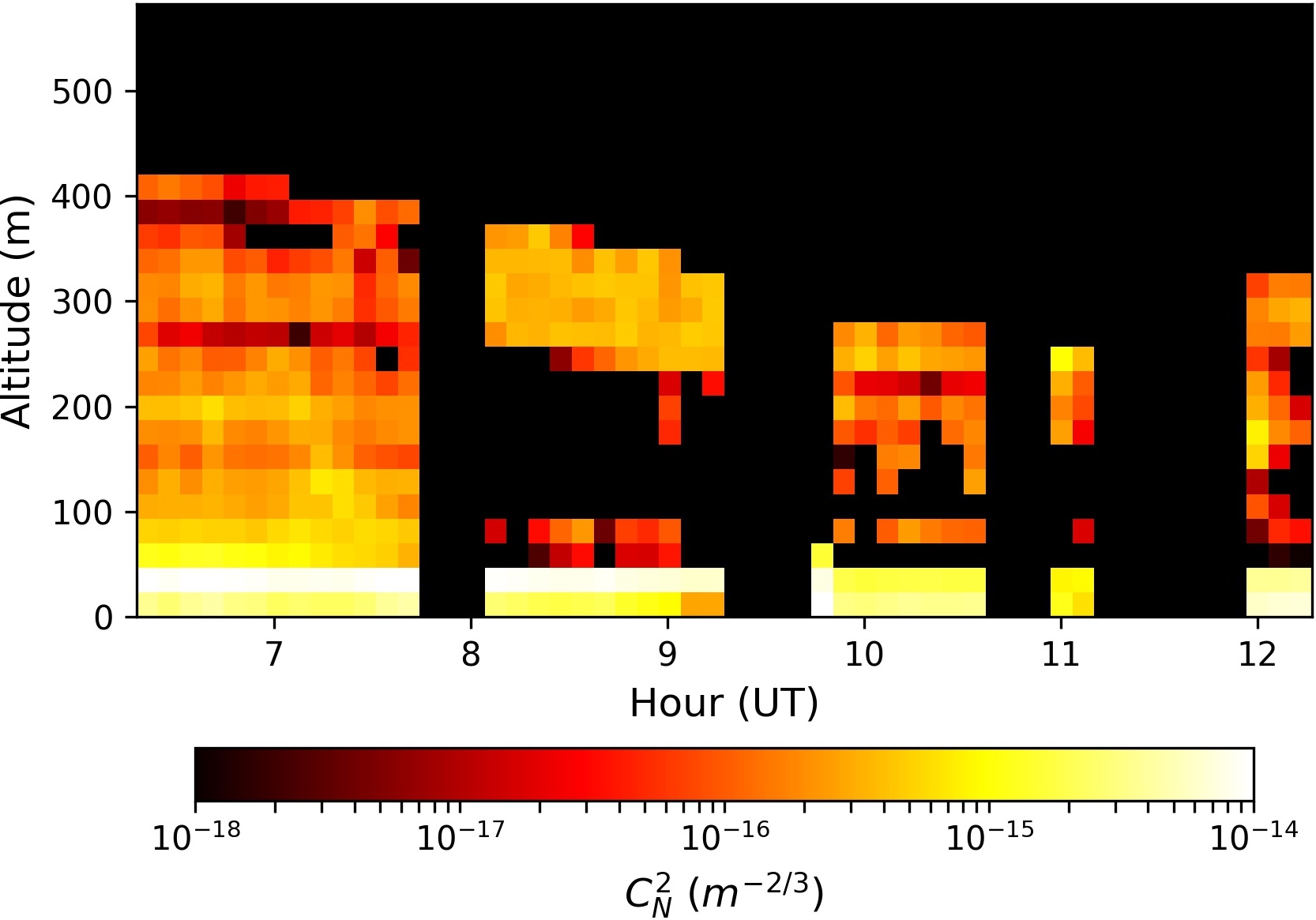}
\caption{Optical turbulence profiles measured with the LOLAS instrument on November 16 (top), 17 (middle) and 18 (bottom) 2013. }
\label{16_11_2013}
\end{figure}

%%%%%%%%%%%%%%%%%%%%%%%%%%%%%%%%%%%%%%%%%%%%%%%%%%

%%%%%%%%%%%%%%%%%%%%%%%%%%%%%%%%%%%%%%%%%%%%%%%%%%
 
\subsection{Wind profiles} 

Each turbulent layer is carried by the predominant wind at the corresponding altitude. Although the turbulent structure suffers a certain temporal decorrelation while driven by the wind, as investigated by \citet{AAC+16}, the degree of correlation between consecutive scintillation frames is high enough to being able to detect peaks in the spatio-temporal covariance of scintillation images. A layer moving at velocity $\vec{v}$ produces a triple peak that lies a distance $\vec{v}\,\Delta t$ from the correlation center. By determining this distance, the layer velocity is obtained. Details of this method are presented in \citet{avi98} and \citet{AVS01}. 

%%%%%%%%%%%%%%%%%%%%%%%%%%%%%%%%%%%%%%%%%%%%%%%%%%

\subsection{Dome and telescope turbulence subtraction} 

\citet{avi98} and \citet{AVS01} explained how optical turbulence inside the telescope dome for the G-SCIDAR and in the telescope tube for LOLAS can be estimated using the {spatio-temporal} covariances. We use their method to remove dome and telescope contributions from all the $C_N^2(h)$ profiles.

\section{Experimental Setup}

 The Observatorio Astron\'omico Nacional in San Pedro  M\'artir (OAN-SPM), operated by the Instituto de Astronom{\'i}a of the Universidad Nacional Aut\'onoma de M\'exico, is situated on the Baja California peninsula at 31$^\circ$ 02' North, 115$^\circ$ 29' West location at an altitude of 2800 m above sea level. It lies within the 
 Northeastern part of the San Pedro M\'artir (SPM) National Park, at the summit of the SPM sierra. { The experimental setup location is presented in Fig.~\ref{maps} which shows an aerial view (top) and a contour map level curves (bottom).}
LOLAS equatorial mount was fixed to a 1-m tall concrete column on a hill that faced a cliff towards the North-East. The G-SCIDAR was adapted to the 2.1-m telescope (2.1mT) which is installed on top of a 20-m tall building. The horizontal distance between both instruments was approximately 100~m {\color{black} and the G-SCIDAR was 23~m higher than the LOLAS. } 

For the G-SCIDAR, images were taken at a rate of 22.8 frames per second on an Andor Luca R Electron Multiplied Charge-Coupled Device (EMCCD) camera of 1004$\,\times\,$1000 pixels that were binned 2$\,\times\,$2. The exposure time of each image was either 1 or 2 ms, depending on the stellar magnitudes and wind conditions. The spatial sampling on the analysis plane was 1 cm. 

For LOLAS we used an Andor iXon EMCCD of 512$,\times\,$512 pixels. The detecting area was set to a window of 512$\,\times\,$160 pixels that were binned 2$\,\times\,$2. The corresponding spatial sampling on the analysis plane was 9.7-mm wide. Images were acquired at a rate of 77 frames per second and an exposure time of 2~ms. The LOLAS instrument operated with its dedicated 16-inch  Ritchey-Chr\'etien telescope. LOLAS data acquisition system {\color{black} and optical re-design} was described by \citet{2016PASP..128j5001A}. The virtual distance $d$ was set to 3000 m for the G-SCIDAR and 1100 m for LOLAS. The number of frames to compute a single autocovariance was set to 3000 and 30000 for the G-SCIDAR and the LOLAS, respectively. {\color{black} LOLAS requires that many frames to compensate its small collective area. As a consequence, one profile is obtained every 6.5 minutes. \citet{2004PASP..116..682A} showed in their Figure 8 that the temporal autocorrelation of the $C_N^2$ values in the first 2 km has a value of 0.9 at 6.5 minutes, indicating that the turbulence strength integrated in that altitude slab does not vary significantly in that time lapse.  Nevertheless, it is not impossible that the high resolution $C_N^2$ distribution close to the ground varies faster than the sampling time of the LOLAS profiles.}

{\color{black} Under those instrumental parameters, in practice, the limiting combined stellar magnitudes for G-SCIDAR and LOLAS were approximately 6.5 and 7, respectively. It is worth recalling that while the G-SCIDAR useful signal decreases when the magnitude difference increases, that of LOLAS is independent of  the magnitude difference, as clearly shown in Eqs.~\ref{Autocov-GS} and \ref{Autocov-LOLAS} through the dependence and independence on the parameter $b$ (defined in Eq.\ref{eq:b}), respectively. The independence of the sensitivity upon the stellar magnitude is also encountered in the Stereo-SCIDAR  \citep{2014MNRAS.437.3568S}}.

Measurement campaigns took place in November 2013, May 2014 and May 2015. In 2013 only LOLAS was operated. While in 2014 both G-SCIDAR and LOLAS worked simultaneously. Finally, in 2015 only  G-SCIDAR measurements were taken. 

Table~\ref{targetsT} summarizes the double star targets used for each instrument during the three campaigns, {where coordinates, magnitudes, separations and position angles for each stellar pair, are presented.}

Table~\ref{observationsT} gives an overview of the data gathered during the three campaigns, which yield a total number of 2423 G-SCIDAR and 164 LOLAS turbulence profiles. 
%%%%%%%%%%%%%%%%%%%%%%%%%%%%%%%%%%%%%%%%%%%%%%%%%%

\begin{figure*}
\centering 
\includegraphics[width=\textwidth]{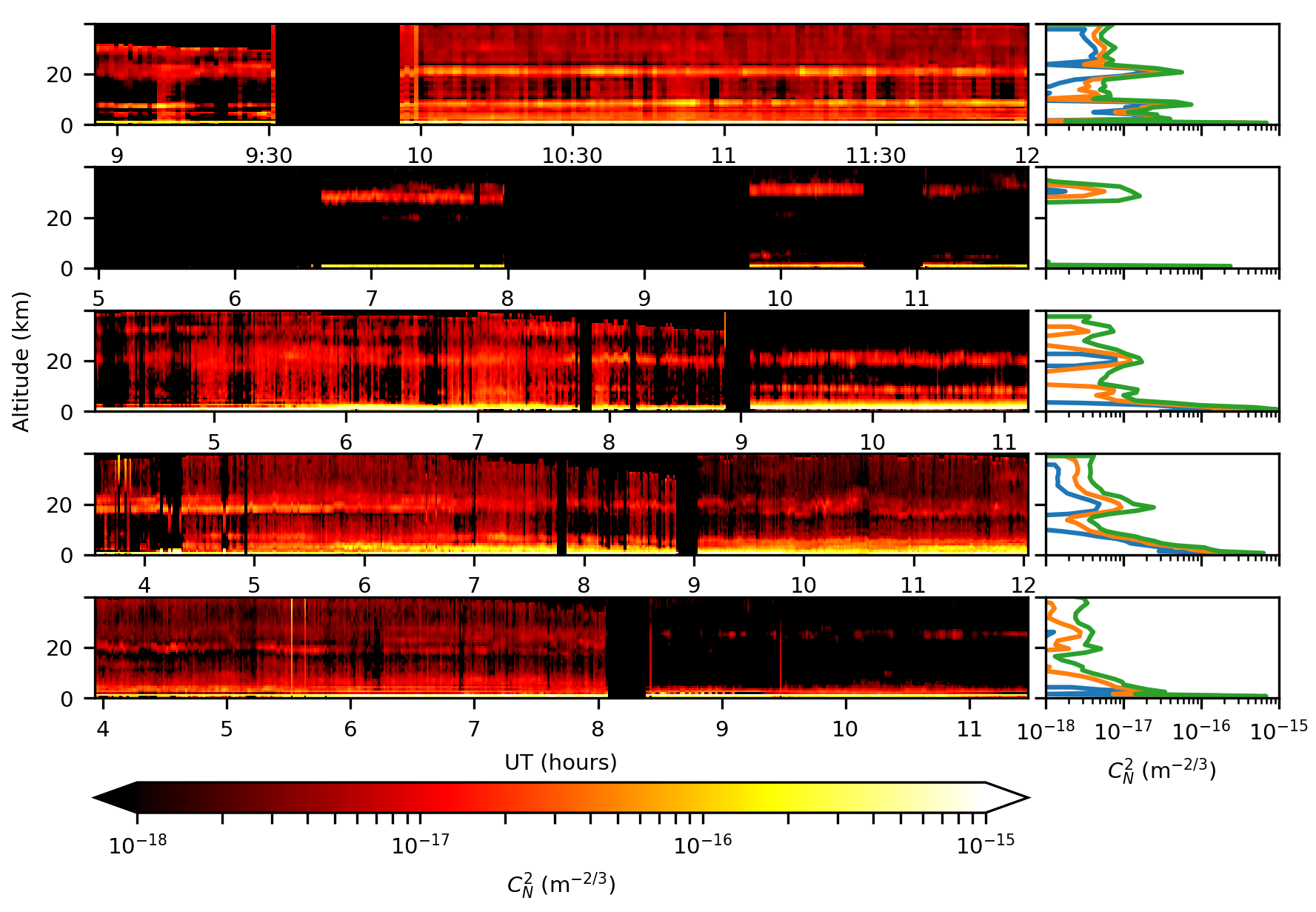}
\caption{Optical turbulence evolution measured with the G-SCIDAR on the campaign of May, 2014. Each row corresponds to one night: May 10, 11, 12, 13 and 14,  from top to bottom. $C_N^2$ values are coded as shown in the color bar on the bottom. The panel on the right shows on a logarithmic scale for each night the median (orange), 1st quartile (blue) and 3rd quartile (green) values of the $C_N^2$ distributions.}
\label{05_2014}
\end{figure*}

%%%%%%%%%%%%%%%%%%%%%%%%%%%%%%%%%%%%%%%%%%%%%%%%%%

%%%%%%%%%%%%%%%%%%%%%%%%%%%%%%%%%%%%%%%%%%%%%%%%%%

\begin{table*}
\begin{center}
%\begin{tabular}{| c | c | c | c | }
\begin{tabular}{| l | l | c | c | r | c | c |}
    
  \hline
  Campaign & Target & R.~A. & Dec.   & Magnitudes & Separation  & P. A. \\
   & & \multicolumn{2}{c}{(J2000)} & (V) & (\arcsec) & ($^\circ$) \\
  
  \hline
  
  2013 & AG+34~259, 260 (15 Tri) & 02h 35m 46.82s &  +34$^\circ$ 41' 15.2'' & 5.4, 6.7 & 142.4 &  16 \\
  
  2013 & 11,\,12 Cam &  05h 06m 08.46s & +58$^\circ$ 58' 20.6'' & 5.1, 6.1 & 181.3  & 10 \\
  
  %2014 & STFA 23AB (32 Com) &   12h 52m 12.26s & +$17^\circ$ 04' 26.2'' &195.9 & 6.5, 6.99 & 50.7 & Coma Berenices\\
  
  2014 & 32, 33 Com &   12h 52m 12.26s & +17$^\circ$ 04' 26.2'' & 6.3, 6.9 & 195.9  & 51 \\
  
  %2014 & STFA 37AD (ES Epsilon 1,\,2 Lyr, ADS 11635 )  &  18h 44m 20.34s & +$39^\circ$ 40' 12.4" & 208.6  & 5.15, 5.38 & 172 & Lyra\\
  
  2014 & 17,\,16 Dra  &  16h 36m 13.72s & +52$^\circ$ 55' 27.8" & 5.4, 5.5 & 89.3  & 193 \\
  
  2014 & eps01,\,eps02 Lyr &  18h 44m 20.34s & +39$^\circ$ 40' 12.4" & 4.7, 5.2 & 208.6   & 172 \\
  
  %2014 & STFA 30AC (MAL)  &  16h 36m 13.72s & +$52^\circ$ 55' 27.8" & 89.3 & 5.38, 5.50 & 193 & Draco\\

  \hline

  2015 & ADS6012AB (19 Lyn) & 07h 22m 52.00s & +55$^\circ$ 16' 53.0" & 5.8, 6.7 &  14.7   & 316 \\
  
  2015 & ADS8108AB & 11h 16m 04.03s & +52$^\circ$ 46' 23.4" & 6.2, 7.4 &  12.4  & 344 \\
  
  2015 & STF1604AB & 12h 09m 28.52s & -11$^\circ$ 51' 25.5" & 6.6, 9.1  &  10.1 & 90 \\
  
 %2015 & STF1604AC & 12h 09m 28.52s & -$11^\circ$ 51' 25.5" &  10.1 & 6.9, 8.1 & 4.6 \\
  
  2014, 2015 & STF1744AB (Mizar A,B) & 13h 23m 55.42s & +54$^\circ$ 55' 31.5" & 2.2, 3.9 &  14.6 & 153 \\
  
  2015 & ADS9728AB  & 15h 38m 40.08s  & -08$^\circ$ 47' 29.4" & 6.5, 6.5 &  11.9  & 190 \\
  
  2014, 2015 & STF2280AB (100 Her A,B) & 18h 07m 49.56s  & +26$^\circ$ 06' 04.4" & 5.8, 5.9 &  14.4 & 183 \\ 
  
  \hline

\end{tabular}
\end{center}
\caption{Double-star targets used for the instruments LOLAS (top 5 rows) and G-SCIDAR (bottom 6 rows). }
\label{targetsT}
\end{table*}

%%%%%%%%%%%%%%%%%%%%%%%%%%%%%%%%%%%%%%%%%%%%%%%%%%

\begin{table}
\begin{center}
\begin{tabular}{| c | c | c |}
  \hline
  Date & G-SCIDAR & LOLAS\\
  (DD/MM/YYYY) & profiles & profiles\\
  \hline
  16/11/2013 & - & 28\\
  17/11/2013 & - & 41\\
  18/11/2013 & - & 37\\

  \hline
  10/05/2014 & 170 & 17\\
  11/05/2014 & 207 & -\\
  12/05/2014 & 445 & 22\\
  13/05/2014 & 542 & 19\\
  14/05/2014 & 475 & -\\
  \hline
  03/05/2015 & 30 & -\\
  04/05/2015 & 90 & -\\
  05/05/2015 & 157 & -\\
  06/05/2015 & 180 & -\\
  07/05/2015 & 127 & -\\
  \hline
  Total & 2423 & 164\\
  \hline

\end{tabular}
\end{center}
\caption{Number of profiles measured each night. }
\label{observationsT}
\end{table}

%%%%%%%%%%%%%%%%%%%%%%%%%%%%%%%%%%%%%%%%%%%%%%%%%%

\section{Results and discussion}
\label{sec:results}

%%%%%%%%%%%%%%%%%%%%%%%%%%%%%%%%%%%%%%%%%%%%%%%%%%

\subsection{Measurements of 2013}

On November 16 2013, we only measured profiles with the LOLAS instrument, using the first two targets of Table~\ref{targetsT}. The upper panel of Fig.~\ref{16_11_2013} shows the optical turbulence evolution of this first night. Noticeably, as the night advanced the turbulence closer to the ground became weaker. This could be due to ground cooling as the night progresses. The maximum altitude resolution attained was 20 metres.  
Measurements of the following two nights are shown in the middle and bottom panels of Fig.~\ref{16_11_2013}. Optical turbulence is strongly concentrated below 50 metres. 

Interestingly, in the first 3 to 4 hours of each night, the optical turbulence does not peak at ground level but between 25 to 50 m above the ground (2nd altitude bin). After that, either the distribution gets homogeneous within the first 50 m or the highest $C_N^2$ values are located at ground level, as one would expect. Optical turbulence occurs where two conditions are met: air must be flowing turbulently (which is favored by wind shears) and the temperature has a sufficiently steep vertical gradient. Wind velocity values obtained from the LOLAS data do not reveal any clear trend: for example, on November 16 the {\color{black} average speed of the strongest-turbulence layer} in the first half of the night was higher than in the second half (15.8 and 9.4 m\,s$^{-1}$, respectively), whereas on November 18 the situation was opposite (3.8 and 5.45 m\,s$^{-1}$, respectively). This leads us to suspect that in the first half of the night there is a steep temperature gradient between 25 to 50 m above the ground that tends to vanish towards the second half of the night. The trees on the site might constitute a factor that tends to homogenize the temperature in the first 25 m of altitude. {\color{black} One could suspect the turbulence  peak between 25 to 50 m to be a consequence of the 2.1mT building wake, however, both nights, the wind was coming from the west-northwest, that is in opposite direction of the 2.1mT.}

From each profile we calculated the corresponding seeing expressed in arcsec as: 

\begin{equation}
\epsilon =1.08\times10^6 \,\lambda^{-1/5}\left[\int_{0}^{h_{max}} \mathrm{d}h C_N^2(h)\right]^{3/5}.
\label{eq:seeing}
\end{equation} 

 The median value of the seeing measured with LOLAS, computed from the 2013 profiles was 0.57 arcsec. This might seem a high number considering that only turbulence below 400 m is considered, but one has to bear in mind that the LOLAS was installed at ground level {\color{black}where} most of the turbulence is generated. This value is in excellent agreement with the median Ground Layer turbulence (defined as turbulence below 500-m) of 0.59 arcsec measured by \citet{2012MNRAS.426..635S}. 

%%%%%%%%%%%%%%%%%%%%%%%%%%%%%%%%%%%%%%%%%%%%%%%%%%

\subsection{Measurements of 2014}

On May 2014 the G-SCIDAR was mounted on the 2.1mT to measure turbulence profiles up to 20 km above the ground with an altitude resolution ranging from 265 to 381 m, depending on the double star used. The LOLAS instrument was operating during three nights together with the G-SCIDAR. We discuss the results of simultaneous measurements during those nights later in this section.

Figure~\ref{05_2014} shows the optical turbulence evolution measured with the G-SCIDAR. All along each night, the turbulent distribution is fairly stable. We detected the jet stream  between 10 and 15 kilometres in almost all nights as well as a turbulent structure below 5 km. Profiles are corrected from dome turbulence. On May 11, the $C_N^2$ profiles consist almost exclusively of two layers: one stuck to the ground and the other at 15 km approximately. This situation does not repeat itself on any other night reported here. On May 12 a diffuse turbulent structure between 7 and 12 km is observed at the beginning of the night, which ends up narrowing to a 1-km thick layer at approximately 7 km above the ground. {\color{black} In the second half of that same night, the strongest turbulence is found at a height of  few hundred meters rather than at ground level. This interesting feature is more deeply discussed in \S~\ref{sec:simultaneous}}. The median seeing values for May 10 to 14 were 0.87, 0.89, 1.06, 0.95 and 0.69 arcsec, respectively. Similarly, the median dome seeing values were 0.27, 0.26, 0.36, 0.15 and 0.3 arcsec, {\color{black} which are consistent with the median dome seeing of 0.31 arcsec found by \citet{2004PASP..116..682A} in the same telescope. Similar values of dome seeing have been measured at other telescopes: \citet{BT18} found values between 0.2 and 0.26 arcsec at the 4-m Blanco telescope and \citet{LWL+19} reported a dome seeing of 0.4 arcsec at the 88-inch telescope of the University of Hawaii in Mauna Kea.}

\begin{figure}
\centering
\includegraphics[width=0.9\columnwidth]{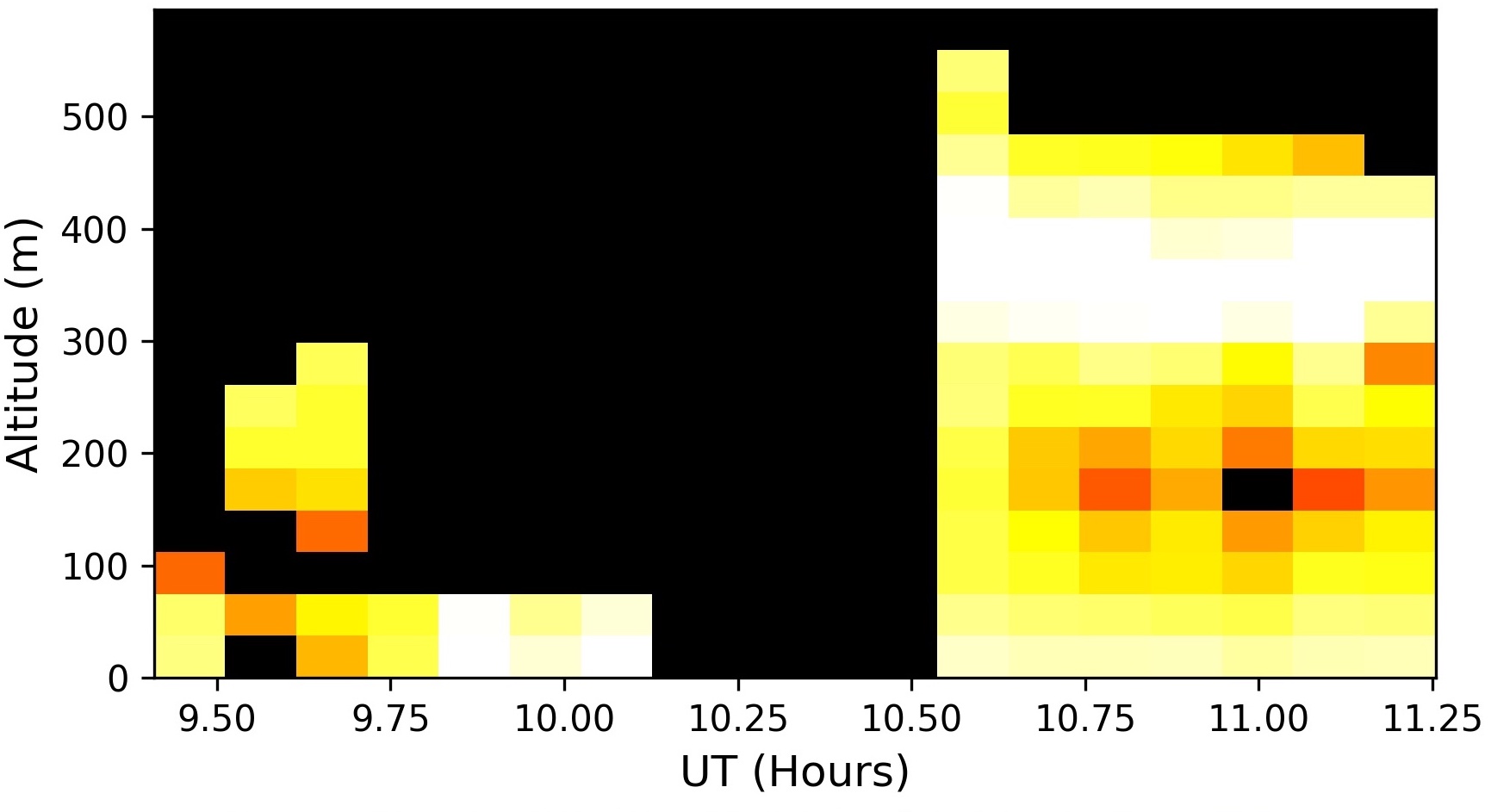}
\includegraphics[width=0.9\columnwidth]{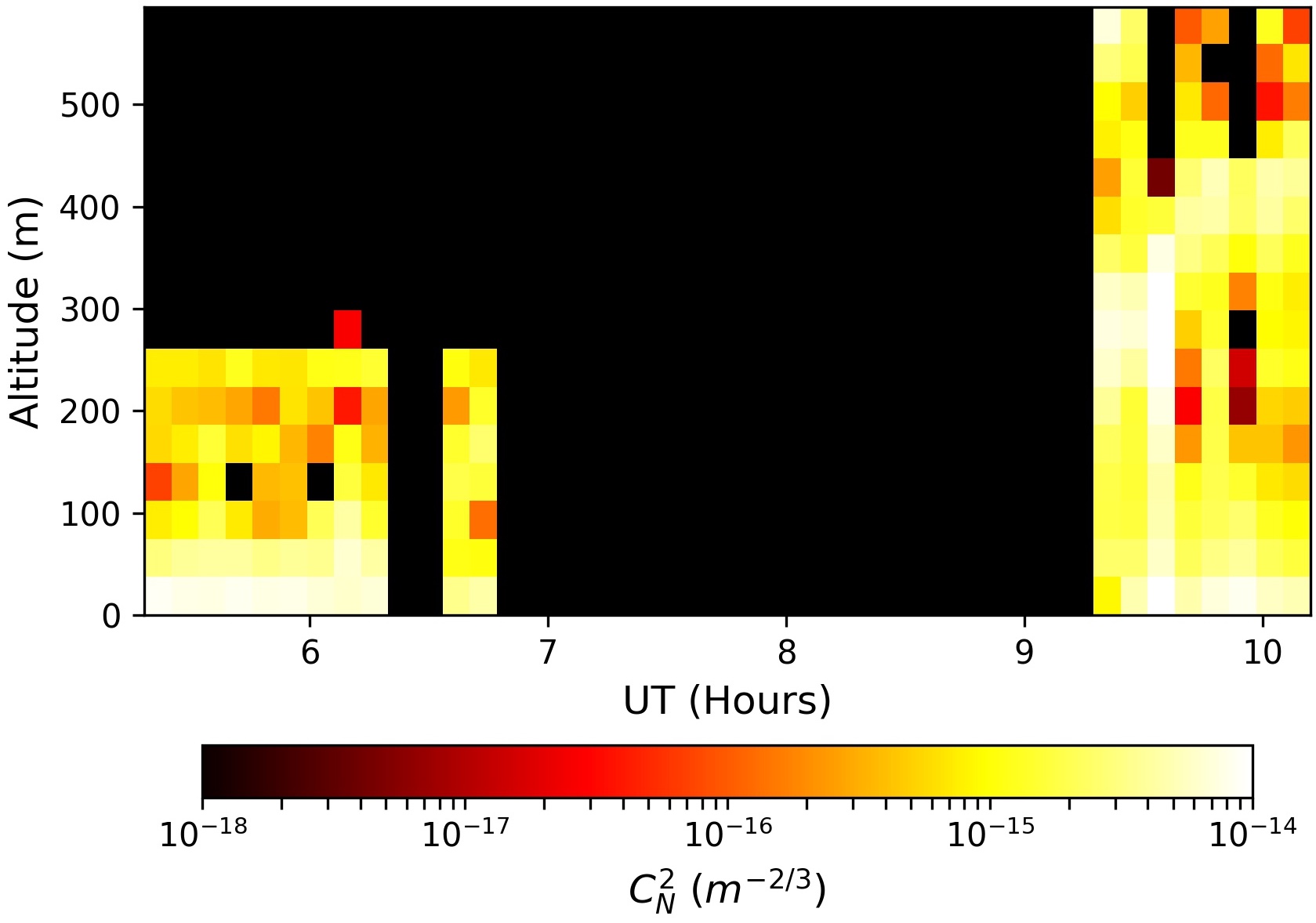}
\caption{Optical turbulence profiles measured with the LOLAS instrument on the nights of May 12 and 13, 2014.}
\label{2014_05LOLAS}
\end{figure}

%%%%%%%%%%%%%%%%%%%%%%%%%%%%%%%%%%%%%%%%%%%%%%%%%%

\begin{figure*}
\centering
\includegraphics[width=\textwidth]{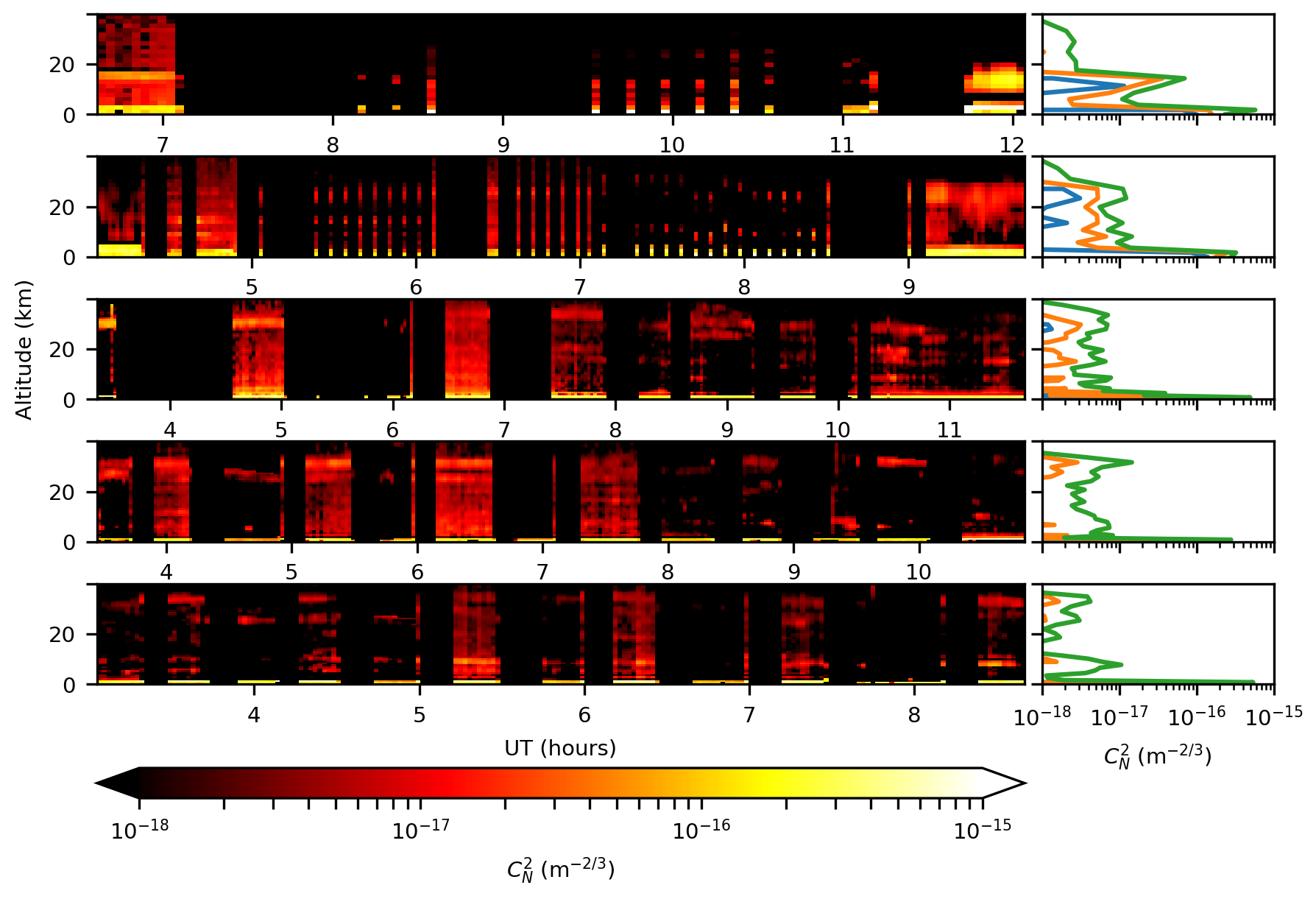}
\caption{Similar to Fig.~\ref{05_2014} but for the nights of 2015, May 3rd to 7th, from top to bottom.}
\label{05_2015}
\end{figure*}

%%%%%%%%%%%%%%%%%%%%%%%%%%%%%%%%%%%%%%%%%%%%%%%%%%

\subsection{Measurements of 2015}
\label{sec:meas2015}

The following year, on May 2015 another campaign was conducted by our team at the OAN-SPM with the G-SCIDAR. Figure~\ref{05_2015} shows the temporal evolution maps of the optical turbulence profiles measured that week. On the first night, two well distinguished predominant layers are detected: one close to the ground and the other at 8 km approximately. The rest of the nights (May 4th to 7th), the high-altitude layer seems much weaker. This situation was also encountered at the site on 2000, May 9 to 14th \citep{2004PASP..116..682A}. It is a very beneficial condition for astronomical observations with adaptive optical compensation since the field of view that can be corrected increases dramatically when high-altitude turbulence is weak. In the 2015 campaign we observed the lowest turbulence among the measurements reported here. The median seeing values obtained were 0.9, 0.7, 0.6, 0.5 and 0.6 arcsec for May 3rd to 7th, 2015, respectively. The corresponding median dome seeing values were 0.5, 0.3, 0.3, 0.2 and 0.2 arcsec. {\color{black}Temporal discontinuities in the data acquisition that can be seen in Fig.~\ref{05_2015} were due to technical problems in the instrument computer. } 

%%%%%%%%%%%%%%%%%%%%%%%%%%%%%%%%%%%%%%%%%%%%%%%%%%

\subsection{Simultaneous LOLAS and G-SCIDAR measurements}
\label{sec:simultaneous}

The simultaneous assessment of low-resolution turbulence profiles in the whole atmosphere and high-resolution profiles in the first few hundred meters enables a profound comprehension of atmospheric optical turbulence and a very complete  site characterization.

\begin{figure}
\centering
\includegraphics[width=\columnwidth]{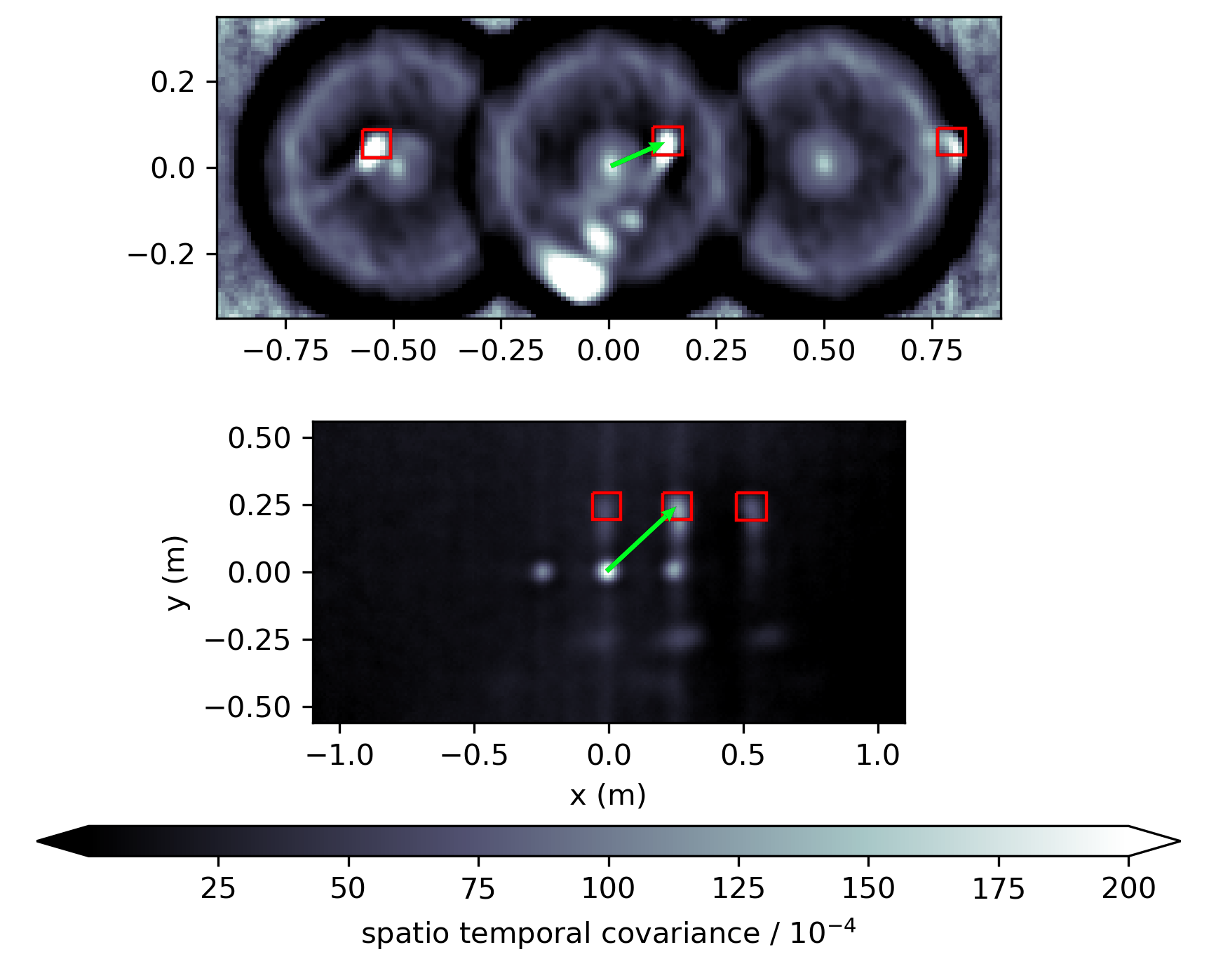}
\caption{
Simultaneous normalised spatio-temporal covariance of scintillation measured at 11:00:44 UT on November 12, 2014. The temporal lag for the temporal correlations was $\Delta t\,=\,$ 13~ms and $\Delta t\,=\,$ 23~ms for LOLAS (top) and G-SCIDAR (bottom). {\color{black}The red squares indicate the correlation triplets that correspond to the strongest layer between 0 and 400 m above the ground. The green arrows indicate the displacement of the central peak from the correlation center, which is produced by the wind velocity.}}
\label{figCross}
\end{figure}

On May 12 2014 we detected a turbulent layer at 350 metres with the LOLAS instrument from 10:30 to 11:30 UT (see Fig.~\ref{2014_05LOLAS}), which we did not see on any other night. This layer and turbulence at the ground cannot be {\color{black} fully resolved as distinct layers on the G-SCIDAR profiles  due to their limited altitude-resolution, but the G-SCIDAR profiles do show that the strongest turbulence is located a few hundred meters above the ground. By} analysing the spatio-temporal covariance functions obtained at the same time with the G-SCIDAR and LOLAS, like those shown in Fig.~\ref{figCross}, we are able to identify that layer in both instruments. In the LOLAS map (top panel of Fig.~\ref{figCross}) the layer at 350 m corresponds to the triplet marked with the red squares. The distance between the central and lateral peaks corresponds to 350 m in altitude. Given the temporal lag between images used to compute the covariance ($\Delta t\,=\,$ 13~ms), the displacement of the central peak with respect to the center of the covariance map of 14 pixels which corresponds to 13.6~cm, and the elevation angle of the star, the traced layer is moving at a horizontal speed of 12 $\mathrm{ms}^{-1}$ approximately. On the G-SCIDAR covariance map (Fig.~\ref{figCross}-bottom), the triplet marked with the red squares corresponds to a layer between 0 and 400 m that moves horizontally at a speed of 14 $\mathrm{ms}^{-1}$ approximately, which is in good agreement with the layer detected with LOLAS. This coincidence gives us confidence on the measurements performed with both instruments. {\color{black} Furthermore, we compared the integral of the LOLAS $C_N^2$ values from 200 to 500 m with $C_N^2(h_1) \delta h$ ($h_1=350$~m) obtained simultaneously with the G-SCIDAR. We find a very good agreement, with a mean relative variation of 4\%.} 

\begin{figure}
\centering
\includegraphics[width=0.4\textwidth,height=0.32\textwidth]{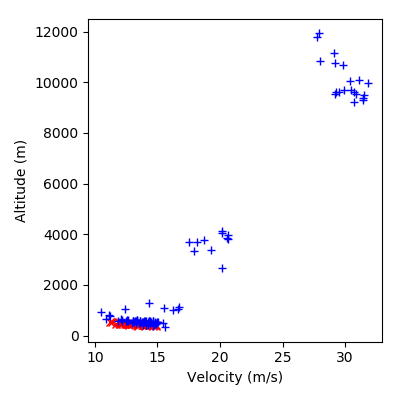} \\
\includegraphics[width=0.4\textwidth,height=0.32\textwidth]{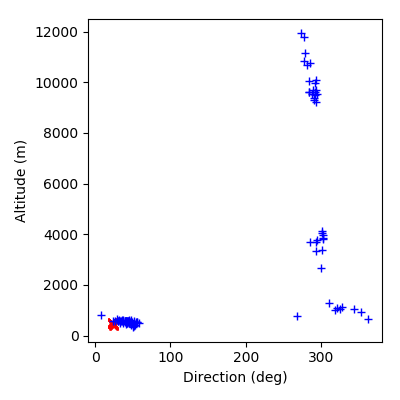}
\caption{Horizontal velocity of optical turbulent layers measured with the LOLAS (red) and G-SCIDAR (blue) instruments simultaneously on May 12, 2014 from 10:30 to 11:30 UT.  On the top is shown the altitude vs. speed of each detected layer. On the bottom is shown the direction from which the layer is moving. Angles $0^{\circ}$ and $90^{\circ}$ correspond to North and  East respectively. }
\label{SimultaneoVel}
\end{figure}

Figure~\ref{SimultaneoVel} shows all velocities of the optical turbulent layers measured with both instruments between 10:30 and 11:30 of May 12 2014 UT. Those velocities are those of the wind at the corresponding altitudes. The wind direction (bottom panel of Fig.~\ref{SimultaneoVel}) is obtained taking into account the position angle of the double star used as light source. The agreement between LOLAS and G-SCIDAR estimates is very good. The wind at low altitude comes predominantly from the North-North-East, presumably carrying warm air from the desert that climbs along the cliff (see Fig.~\ref{maps}) and interacts with colder air above the observatory, producing this very strong optical turbulence at 350 m above the ground.

\subsection{Historical G-SCIDAR statistics at OAN-SPM}

 Turbulence profiles have been measured with G-SCIDAR at the OAN-SPM in 1997, 2000  \citep[see][]{1998PASP..110.1106A,2004PASP..116..682A,2006PASP..118..503A,2007RMxAC..31...71A}, 2014 and 2015. The latter two campaigns are reported here. Some similarities in the vertical distribution of $C_N^2$ obtained in different years can be highlighted: Turbulence close to the ground is almost always the strongest. Many times a layer is found between 8 to 15 km above the ground, which can be attributed to the jet stream. Some times a layer between 5 and 7 km is also noticed. On several nights, turbulence in most of the atmosphere above three kilometers is very low, as discussed in $\S~\ref{sec:meas2015}$.

\begin{figure}
\centering
\includegraphics[width=0.45\textwidth]{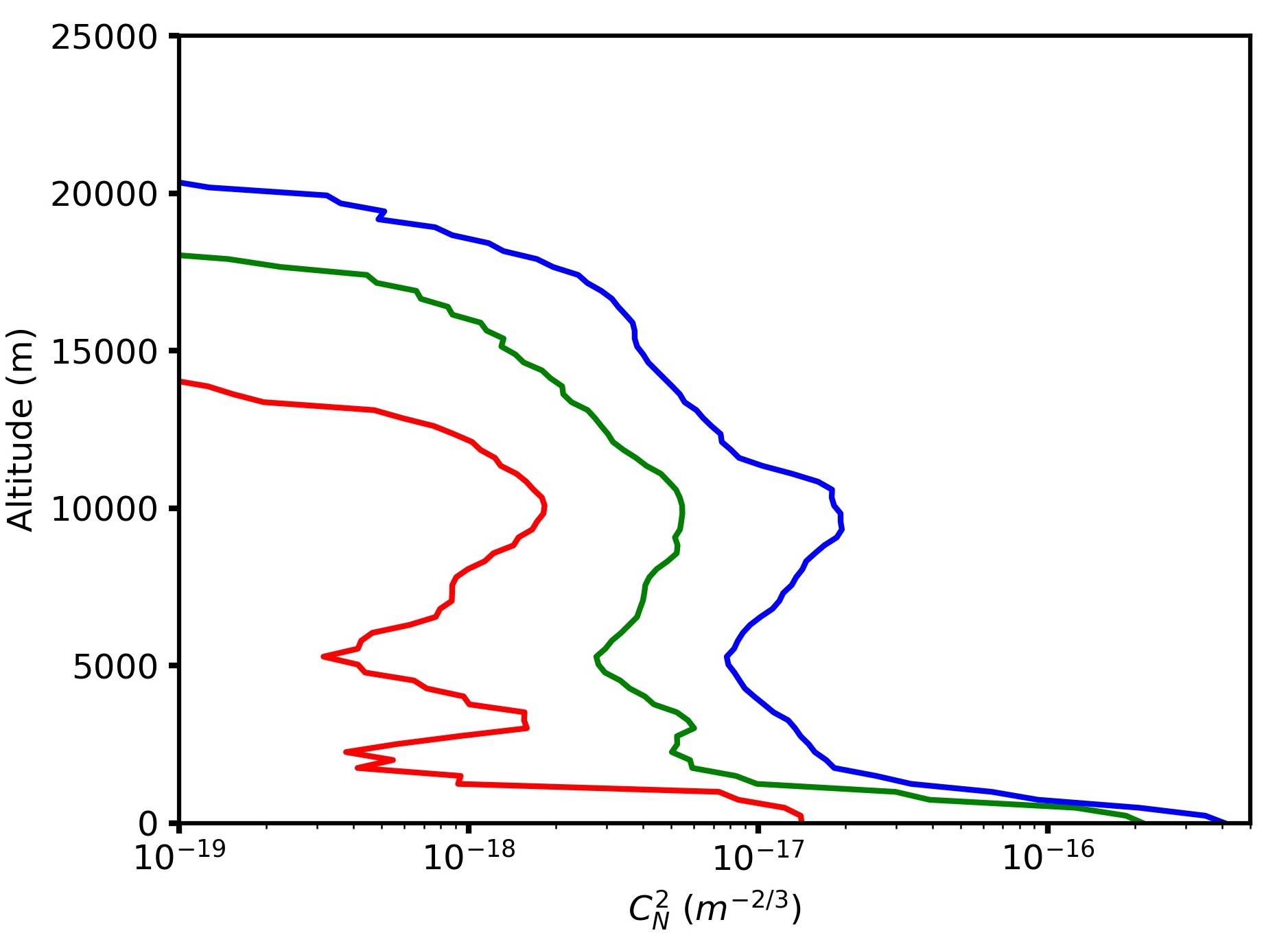}
\includegraphics[width=0.45\textwidth]{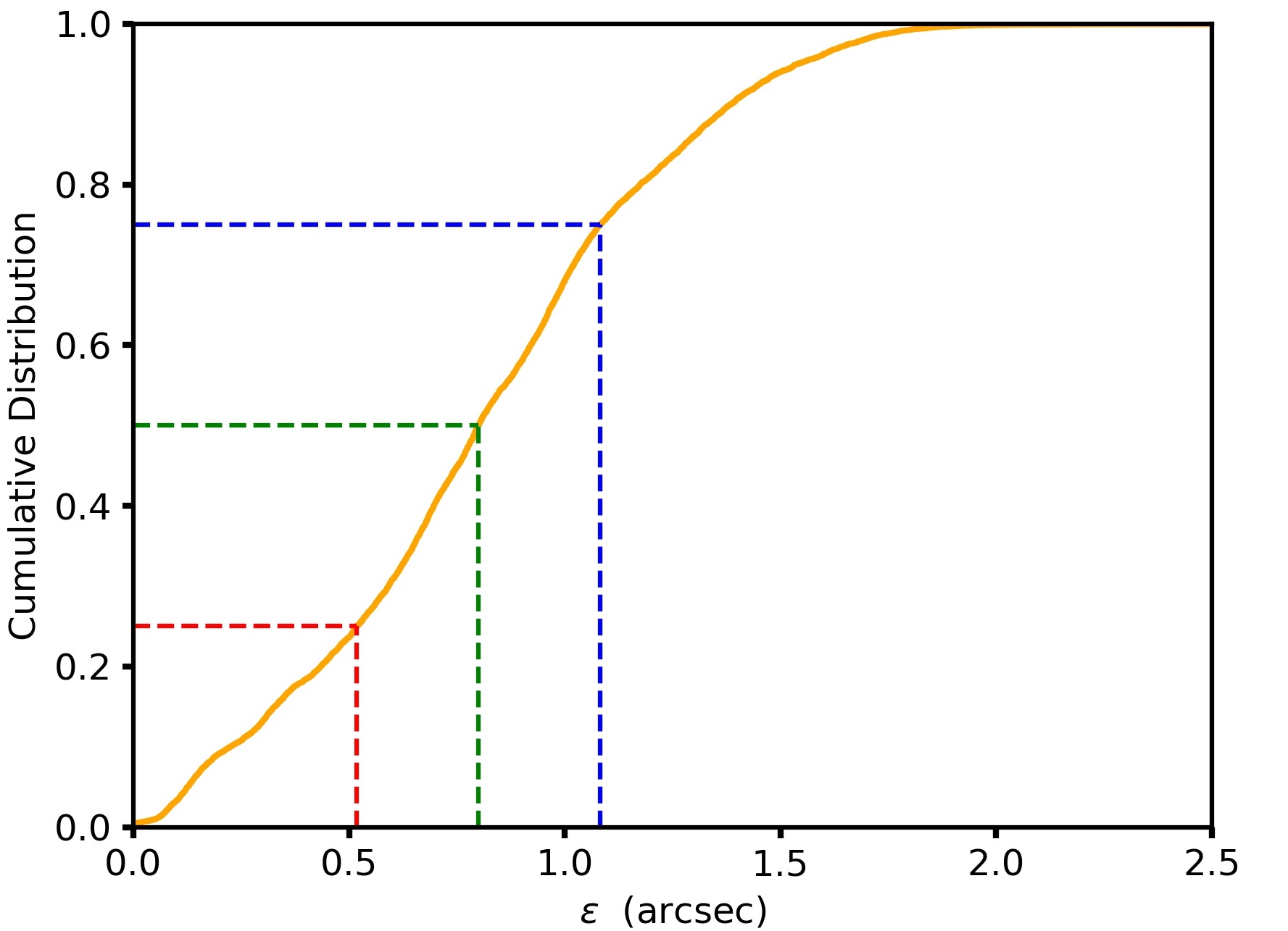}
\caption{Statistics of all $C_N^2$ profiles that have been measured with the G-SCIDAR technique at the OAN-SPM in 1997, 2000, 2014 and 2015 campaigns. At the top are shown the median (green), 1st (red) and 3rd quartile profiles. The cumulative distribution of seeing calculated from individual profiles is shown at the bottom. Each profile is corrected from dome turbulence. Green, red and blue lines indicate the median, 1st and 3rd quartiles values of the seeing: 0.79, 0.51 and 1.08 arcsec, respectively.}
\label{statall} 
\end{figure}

Figure~\ref{statall} shows statistics of the 7891 G-SCIDAR profiles obtained so far at the OAN-SPM. Every profile is corrected for dome seeing and properly calibrated following \citet{Avila:09}. The {\color{black} top} panel of Fig.~\ref{statall} shows the median, 1st and 3rd quartiles profiles. We note the well defined layer at 10 km and in the 1st and 2d quartiles a peak can be seen at 3 km. On the {\color{black} bottom} of Fig.~\ref{statall} the  cumulative distribution function of the seeing is shown. First quartile, median, and 3rd quartile values are 0.51, 0.79 and 1.08 arcsec, respectively.  

Those values are closely consistent with the seeing measurements performed by the Thirty Meter Telescope project site survey \citep{2009PASP..121.1151S} from 2004 to 2008 at the OAN-SPM with a differential image motion monitor (DIMM). Using this DIMM data, \citet{2012MNRAS.426..635S} reported that the first quartile, median, and 3rd quartile values of the seeing measured during the spring of those years were 0.61, 0.78 and 1.07 arcsec, respectively. 

%%%%%%%%%%%%%%%%%%%%%%%%%%%%%%%%%%%%%%%%%%%%%%%%%%
%%%%%%%%%%%%%%%%%%%%%%%%%%%%%%%%%%%%%%%%%%%%%%%%%%
%%%%%%%%%%%%%%%%%%%%%%%%%%%%%%%%%%%%%%%%%%%%%%%%%%

%%%%%%%%%%%%%%%%%%%%%%%%%%%%%%%%%%%%%%%%%%%%%%%%%%
%%%%%%%%%%%%%%%%%%%%%%%%%%%%%%%%%%%%%%%%%%%%%%%%%%

%%%%%%%%%%%%%%%%%%%%%%%%%%%%%%%%%%%%%%%%%%%%%%%%%%

\section{Conclusions}

Turbulence profiles measured at the OAN-SPM in the whole atmosphere in 2014 and 2015 confirm the results obtained in 1997 and 2000 with the same technique. The seeing statistics computed with all the G-SCIDAR profiles (1997, 2000, 2014 and 2015 campaign) available at the site have a median of 0.79, first and third quartiles of 0.51 and 1.08 arcsec, which confirm the excellent conditions measured at the site with a DIMM during a 4-year measurement campaign. 

High altitude-resolution profiles obtained close to the ground with the LOLAS have provided two important pieces of evidence. On the one hand, optical turbulence can have its maximum between 25 to 50 m and not as close to the ground as generally expected. This might be produced by the presence of trees on the site. And on the other hand, wind blowing from the cliff located towards the north-northeast can produce a very strong turbulence at 300 m above the site. This layer at 300 m is located by the G-SCIDAR as turbulence {\color{black} a few hundred meters above the ground}. The high altitude-resolution of the LOLAS enabled the determination of its real altitude. These results were achieved only because of simultaneous observations with both instruments.

For the development of ground layer or multiconjugate adaptive optics, the precise determination of the turbulence altitude close to the ground is of capital importance. Longer term LOLAS measurement campaigns at the OAN-SPM would be beneficial for the design of future modern telescopes at the site.

%%%%%%%%%%%%%%%%%%%%%%%%%%%%%%%%%%%%%%%%%%%%%%%%%%

\section*{Acknowledgements}

We are deeply grateful to the technical and administrative staff of the (OAN-SPM) for their valuable help in all the logistics and requirements for the observations. {\color{black} We thank the referee Tim Butterley  for a careful revision of the manuscript and his valuable comments.}. Topographic data was kindly provided by R.~G\'omez Mart{\'i}nez and R.~S\'anchez Garc{\'i}a from the Instituto de Ingenier\'ia-UNAM. Financial support for these projects was provided by DGAPA-UNAM through grants IN-103913, IN-102517 and IN-115013. Instruments were developed with CONACYT (Mexico) grants J32412-E and 58291, while further funding was provided by DGAPA-UNAM grants IN-111403 and IN-112606-2.

\bigskip\bigskip\bigskip\bigskip



\bibliographystyle{mnras}
\bibliography{spm_biblio} % if your bibtex file is called example.bib

%%%%%%%%%%%%%%%%% APPENDICES %%%%%%%%%%%%%%%%%%%%%

%\appendix


%%%%%%%%%%%%%%%%%%%%%%%%%%%%%%%%%%%%%%%%%%%%%%%%%%

% Don't change these lines
\bsp	% typesetting comment
\label{lastpage}
\end{document}